\documentclass[epj]{svjour}

\usepackage{graphics}

\begin{document}

\title{On the distinguishability of histograms}  

\author{S. Bityukov\inst{1,2}, N. Krasnikov\inst{2,4}, A. Nikitenko\inst{3}  
\and V. Smirnova\inst{1} 
}

\institute{Institute for high energy physics,  142281 Protvino, Russia \and 
Institute for nuclear research RAS, 117312 Moscow, Russia \and Imperial College, 
London, United Kingdom,  on leave from ITEP, Moscow, Russia \and 
Joint institute for nuclear research, Dubna, Russia }


\abstract{We consider an approach for testing the hypothesis that two 
realizations of the random variables in the form of histograms are taken 
from the same statistical population (i.e.  two histograms are drawn from the 
same distribution). The approach is based on the notion ``significance of deviation''.  
This approach allows to estimate the statistical difference between two 
histograms  using  multi-dimensional test statistics. The distinguishability 
of histograms is estimated with the help  of the construction a number of clones (rehistograms) 
of the observed histograms.}

\maketitle
%


\section{Introduction}\label{sec:introduce} 

The test of the hypothesis that two  histograms are drawn from 
the same distribution is an important goal in many applications. 
For example, this task exists for the  monitoring of the experimental equipment 
in  particle physics experiments. 
Let the experimental facility register the flow of events during two independent time 
intervals $[t_1,t_2]$ and $[t_3,t_4]$. 
Events from first time interval belong to statistical population of events $G_1$, 
events from second time interval belong to statistical population of events $G_2$. 
If facility (beam, detectors, data acquisition system, ...) is in norm during both time intervals 
then the properties of events, registered in the facility during time interval $[t_1,t_2]$, 
is the same as the properties of events, 
registered in the facility during time interval $[t_3,t_4]$, i.e. $G_1 = G_2$. 
If facility is out of norm during one of time intervals then the properties 
of events from statistical population $G_1$ differ from the properties 
of events from statistical population $G_2$, i.e. $G_1\ne G_2$. 
Often the monitoring of the experimental facility is performed with the use of the comparison 
of histograms, which reflect the properties of events.   

Several approaches to formalize and resolve this problem were considered~\cite{Porter}. 
Recently, the comparison of weighted histograms was developed in paper~\cite{Gagun}. 
Usually, one-dimensional test statistics is  used for the comparison of histograms. 
 
In this paper we propose a method which allows to estimate the value of statistical 
difference between histograms with the use of several test statistics. 
As example, we consider the case of two test statistics, i.e. bidimensional test statistic. 

\section{Distribution of test statistics}

Suppose, there are two histograms $hist_1$ and $hist_2$ (with $M$ bins in each histogram) 
as a result of the treatment of two independent samples of events. 
The first histogram is a set of 2M numbers 

$hist_1:~\hat n_{11}\pm \hat\sigma_{11},\hat n_{21}\pm \hat\sigma_{21},\dots,\hat n_{M1}\pm \hat\sigma_{M1}$ 

\noindent
and the second histogram, correspondingly, is a set of 2M numbers also

$hist_2:~\hat n_{12}\pm \hat\sigma_{12},\hat n_{22}\pm \hat\sigma_{22},\dots,\hat n_{M2}\pm \hat\sigma_{M2}$.  

\noindent
The volume of the first sample is $N_1$, i.e. $\displaystyle N_1 \equiv \sum_{i=1}^M \hat n_{i1}$ and 
the volume of the second sample is $N_2$, i.e. $\displaystyle N_2 \equiv \sum_{i=1}^M \hat n_{i2}$. 

The most of methods for the histograms comparison  use single test statistic as a ``distance measure'' 
 for the consistency of two samples of events (see, for example~\cite{Porter}). 

We propose~\footnote{Some details are in ref.~\cite{arXive}.} 
to use  test statistics $\hat S_i,~~i=1,...,M$ (significances of deviation) for each bin 
for the   histograms comparison. 
In the case of two observed histograms we consider the significance of deviation of the following type:  
\begin{equation}
\hat S_i = \displaystyle 
\frac{\hat n_{i1} - K \hat n_{i2}}{\sqrt{\hat \sigma^2_{i1} + K^2 \hat \sigma^2_{i2}}} \,.     
\label{eq:1}
\end{equation}
\noindent
Here  $K = \displaystyle \frac{N_1}{N_2}$ is a coefficient of the 
normalization~\footnote{This coefficient characterizes the ratio of integral characteristics  
of samples under comparison. It may be, for example, the ratio of volumes (in our consideration) 
or the ratio of time intervals for data acquisition of samples.}. 

We use two first  statistical moments  $\displaystyle \bar S = \frac{\sum_{i=1}^M{\hat S_i}}{M}$, and 
$\displaystyle RMS = \sqrt{\frac{\sum_{i=1}^M{(\hat S_i - \bar S)^2}}{M}}$.    
If condition $G_1 = G_2$ ($G_1$ and $G_2$ are taken from the same 
flow of events) takes place then test statistics ($\hat S_i,~~i=1,...,M$) 
obey the distribution which close to the standard normal 
distribution ${\cal N}(0,1)$. 
Correspondingly, the distribution of these test statistics is close to standard normal 
distribution too. In this case our bidimensional test statistic 
(``distance measure between two observed histograms'') 
$SRMS=(\bar S,RMS)$ has a clear interpretation: 
\begin{itemize}
\item if $SRMS=(0,0)$ then histograms are identical; 
\item if $SRMS \approx (0,1)$ then $G_1 = G_2$ (if $\bar S \approx 0$ and $RMS<1$ then 
 the overlapping exists between samples); 
\item if previous relations are not valid  then $G_1 \ne G_2$. 
\end{itemize}

Note that  the relation 

\begin{equation}
\displaystyle RMS^2 = \frac{\hat \chi^2}{M} - \bar S^2\,,  
\label{eq:2}
\end{equation}
\begin{equation}
\hat \chi^2 = \sum_{i=1}^M{\hat S_i^2} \,
\end{equation}
\noindent
shows that 
test statistic $\hat \chi^2$ is a combination  
of two test statistics $RMS$ and $\bar S$.  

\section{Rehistogramming} 

An accuracy of the estimation of statistical moments depends on the number of bins $M$ in histograms  
and observed values in bins. The accuracy can be estimated via Monte Carlo experiments. 
Two models of the statistical populations (pseudo populations) can be produced. 
Each of models represents one of the histograms. 

In considered below example for each of histograms we produced  4999 clones 
by the Monte Carlo simulation  for each bin $i$ of histogram $k$  
using the normal distribution  ${\cal N}(\hat n_{ik},\hat\sigma_{ik}),~i=1,...,M,~k=1,2$.      
As a result there are 5000 pairs of histograms for comparisons. 
The comparison is performed for each pair of histograms (5000 comparisons in our example). 
The distribution of the significances $\hat S_i$ is obtained as a result of  
each comparison. The moments of this distribution are calculated  (in our case $\bar S$ and $RMS$). 
It allows to estimate the errors in determination of statistical moments.

This procedure can be named as ``rehistogramming'' in analogy with 
``resampling'' in the bootstrap method~\cite{Efron}. 

\section{Distinguishability of histograms}

The estimation of the distinguishability of histograms is performed 
with the use  of hypotheses testing.  
``A probability of correct decision'' ($1 - \tilde\kappa$)  about distinguishability of 
hypotheses~\cite{NIM534} is used as a measure of the potential 
in distinguishing of two flows of events ($G_1$ and $G_2$) via comparison of histograms 
($hist_1$ and $hist_2$). 

It is a probability of the correct choice between two hypotheses 
``the histograms are produced by the treatment of events from the same event 
flow (the same statistical population)'' 
or ``the histograms are produced by the treatment of events from different event flows''.  
The value $1 - \tilde\kappa$
 characterizes the distinguishability of two histograms.

For $1 - \tilde\kappa = 1$ the distinguishability of histograms  is 100\%, i.e. 
histograms are produced by the treatment of events from different event flows.    

For $1 - \tilde\kappa = 0$  we can't  distinguish the histograms, i.e. 
histograms are produced  from the same event flow.    

The probability of correct decision $1 - \tilde\kappa$ is a function of 
type I error ($\alpha $) and the type II error ($\beta $)  
testing, namely~\footnote{The type I error $ \alpha $  is the probability to accept the 
 alternative hypothesis if the main hypothesis is correct. 
The type II error $\beta$  is the probability to accept the main 
hypothesis if the alternative hypothesis is correct. Note, the critical region 
(critical value or critical line) in this consideration  
must be chosen correctly, i.e. $\alpha+\beta\le 1$.} 

\begin{equation}
1 - \displaystyle \tilde\kappa = 1 - \frac{\alpha+\beta}{2 - (\alpha+\beta)}\,. 
\label{eq:3}
\end{equation} 

\section{Example}\label{sec:example}

Let us consider a simple model with two histograms  in which  the random variable in each bin obeys 
the normal distribution  
$\displaystyle 
\varphi(x_{ik}|n_{ik})= \frac{1}{\sqrt{2\pi} \sigma_{ik}}~e^{-\frac{(x_{ik}-n_{ik})^2}{2 \sigma^2_{ik}}}\,.$
\noindent
Here the expected value in the bin $i$ is equal to $n_{ik}$ (in this example $n_{i1}=i$) 
and the variance $\sigma^2_{ik}$ 
is also equal to  $n_{ik}$. $k$ is the histogram number ($k=1,2$). This model can be considered as 
the approximation of Poisson distribution by normal distribution. 

All calculations, Monte Carlo experiments and histograms presentation in this paper are performed 
using ROOT code~\cite{ROOT}.  Histograms are obtained from independent samples. 

The example with histograms produced from the same events flow during 
unequal independent time ranges (Fig.~1) shows that the standard deviation 
of the distribution in the picture (right, down) can be used as an estimator of the statistical 
difference between histograms (this distribution is close to $\cal N$(0,1)).

\begin{figure}[htbp]
\begin{center}
           \resizebox{8.2cm}{!}{\includegraphics{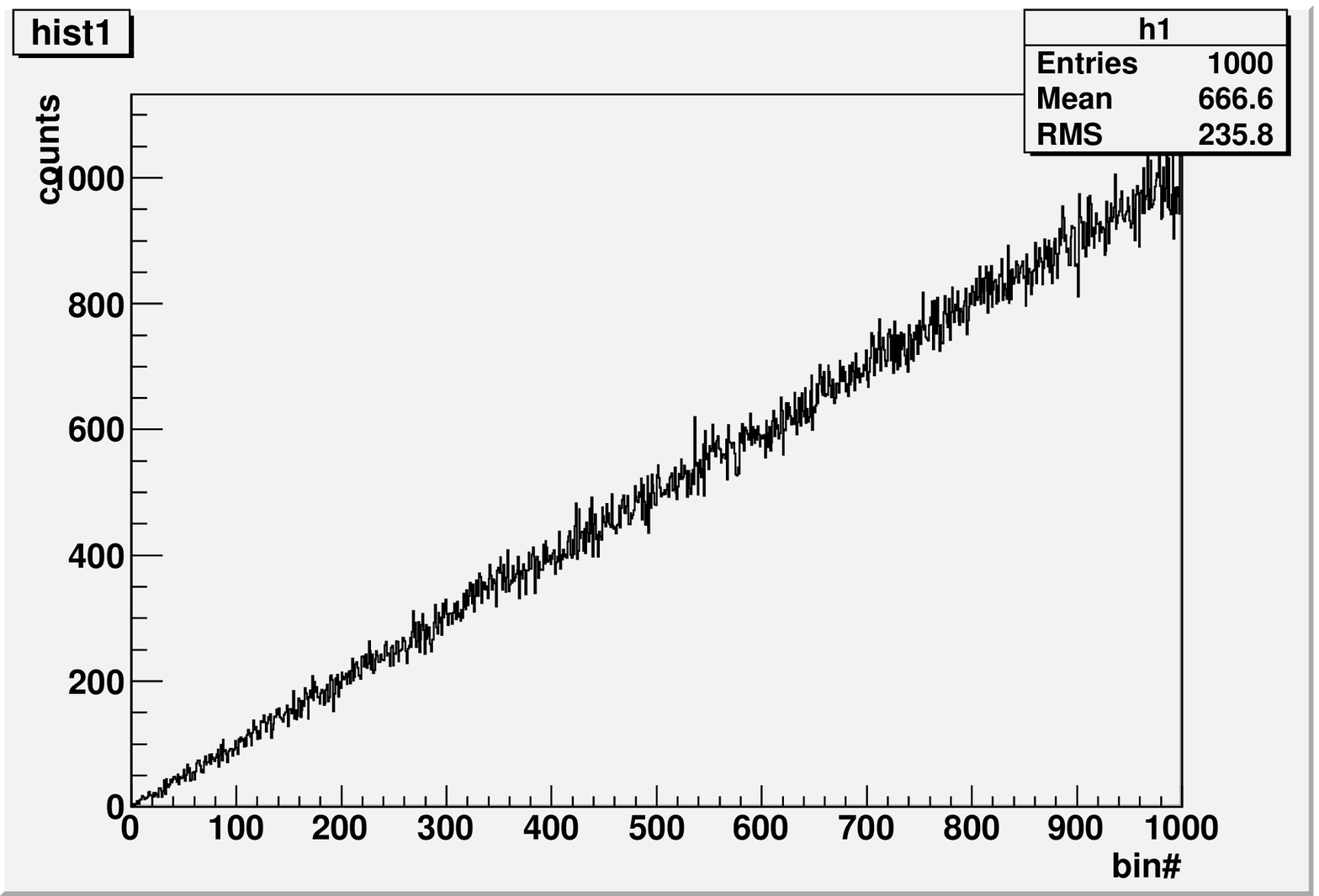}} 
           \resizebox{8.2cm}{!}{\includegraphics{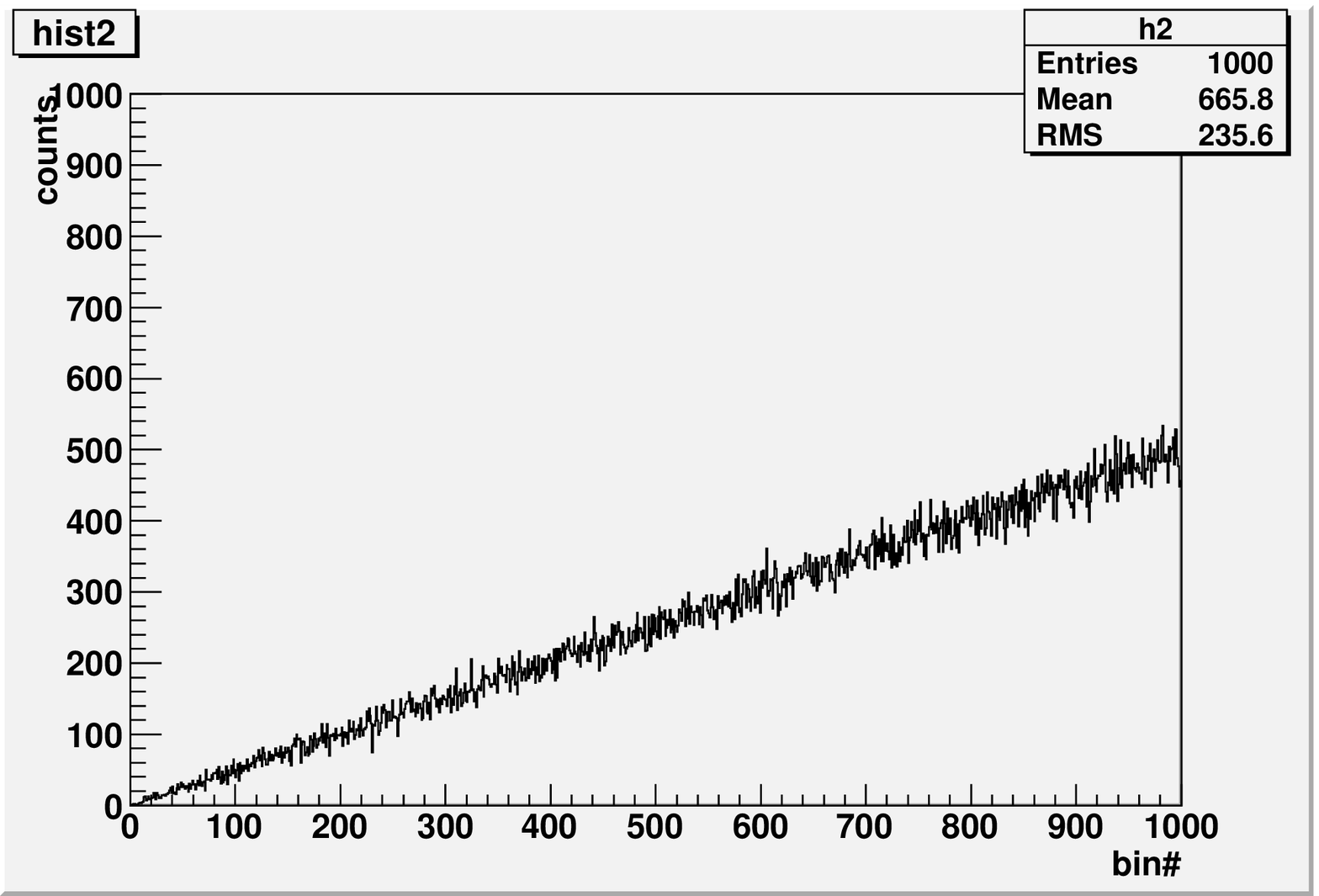}} 
           \resizebox{8.2cm}{!}{\includegraphics{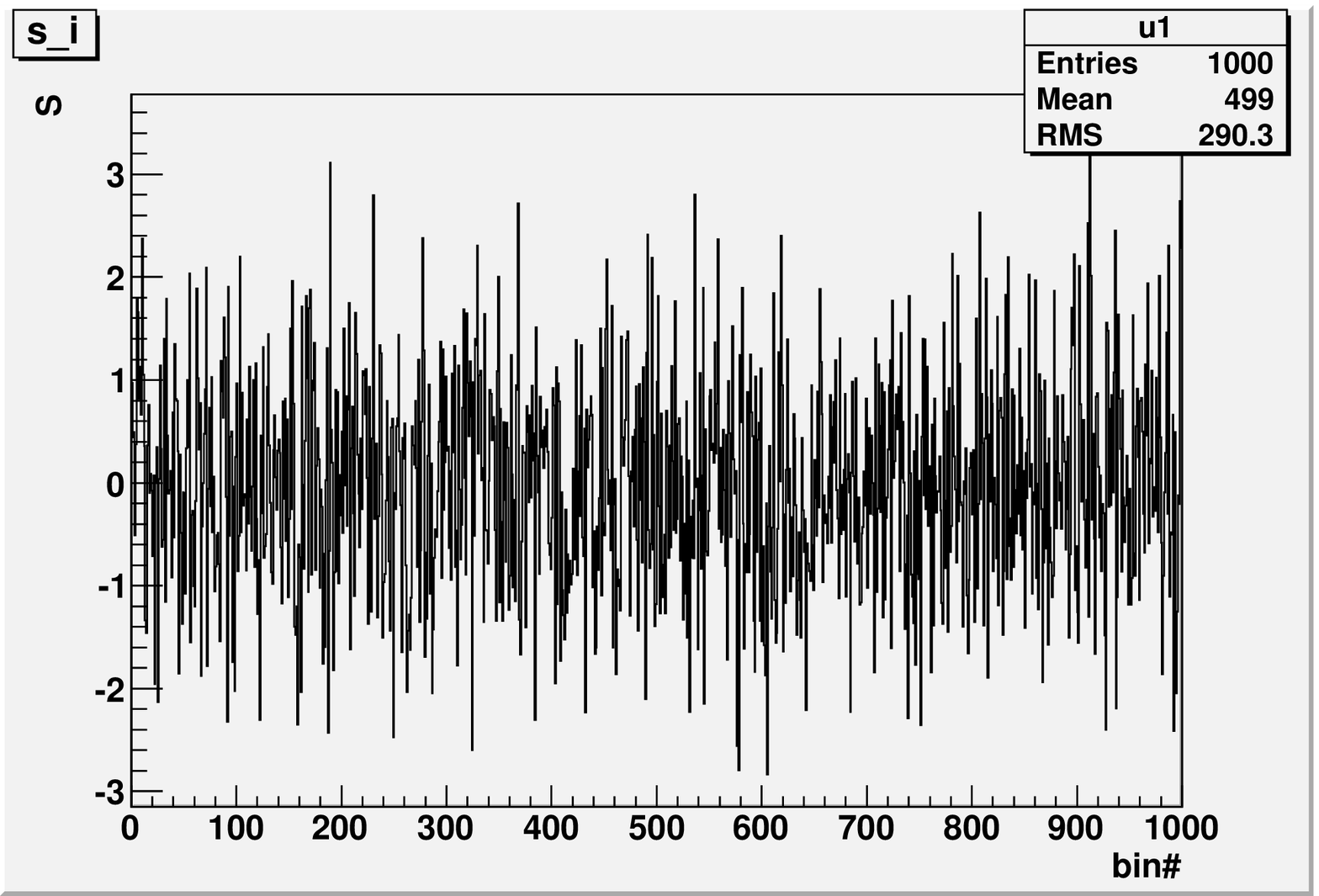}} 
           \resizebox{8.2cm}{!}{\includegraphics{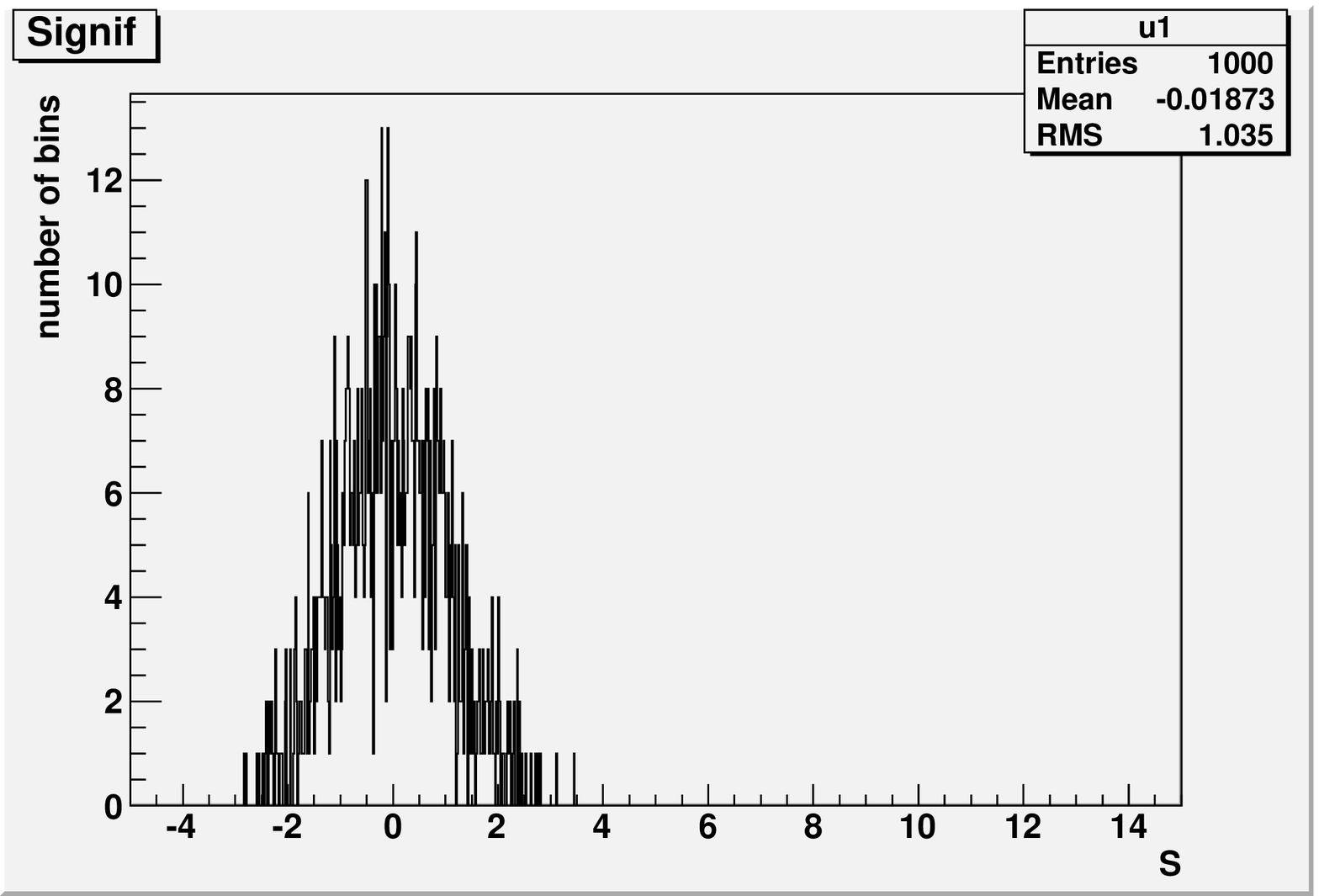}} 
\caption{Triangle distributions in histograms ($M=1000$, $K=2$): 
the observed values $\hat x_{i1}$ in the first histogram (left, up), 
the observed values $\hat x_{i2}$ in the second histogram (right, up), 
observed normalized significances $\hat S_{i}$ bin-by-bin (left, down)
and the distribution of observed normalized significances (right, down).}
\end{center}
\label{fig:1} 
\end{figure}

\begin{figure}
\begin{center}
           \resizebox{8.2cm}{!}{\includegraphics{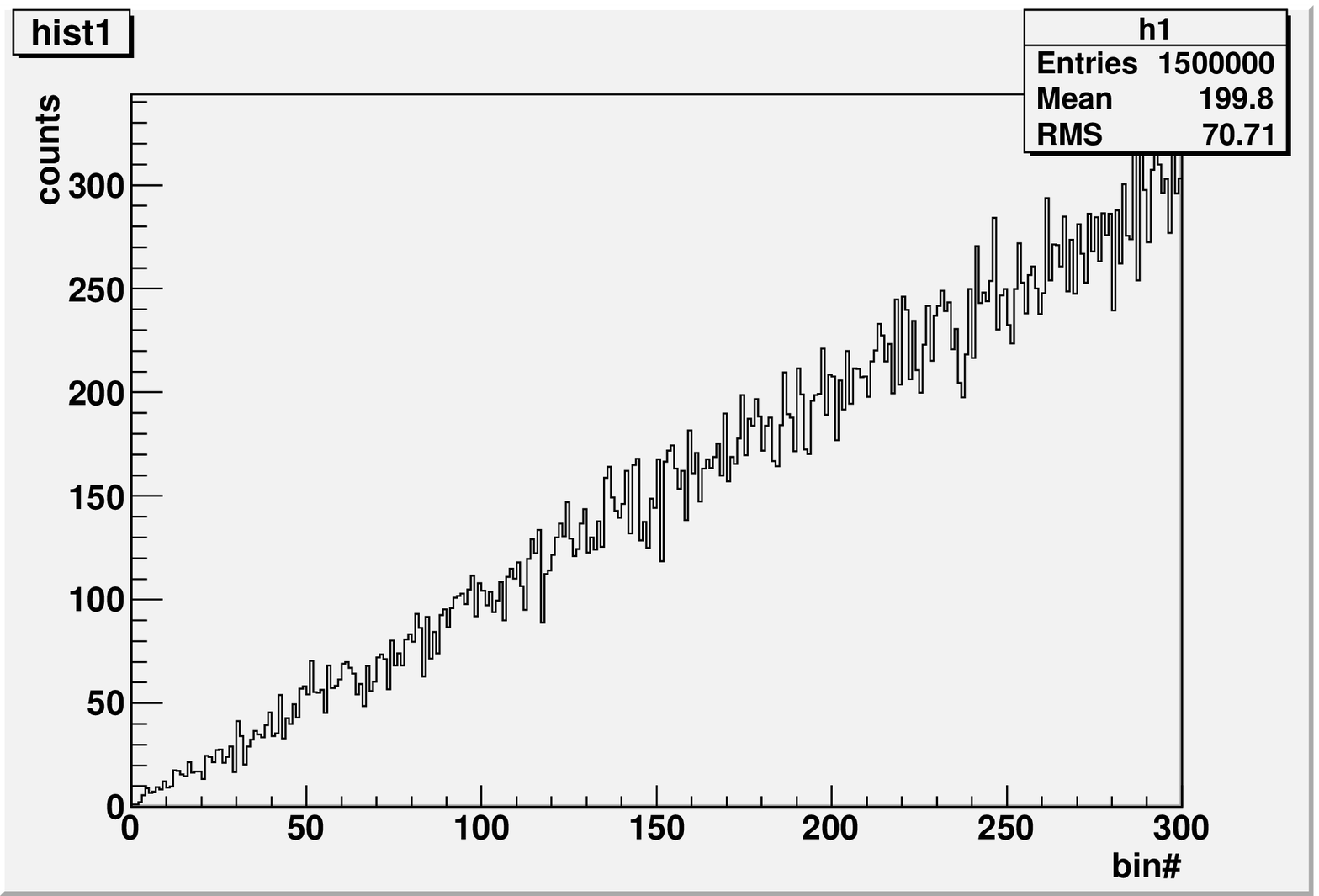}} 
           \resizebox{8.2cm}{!}{\includegraphics{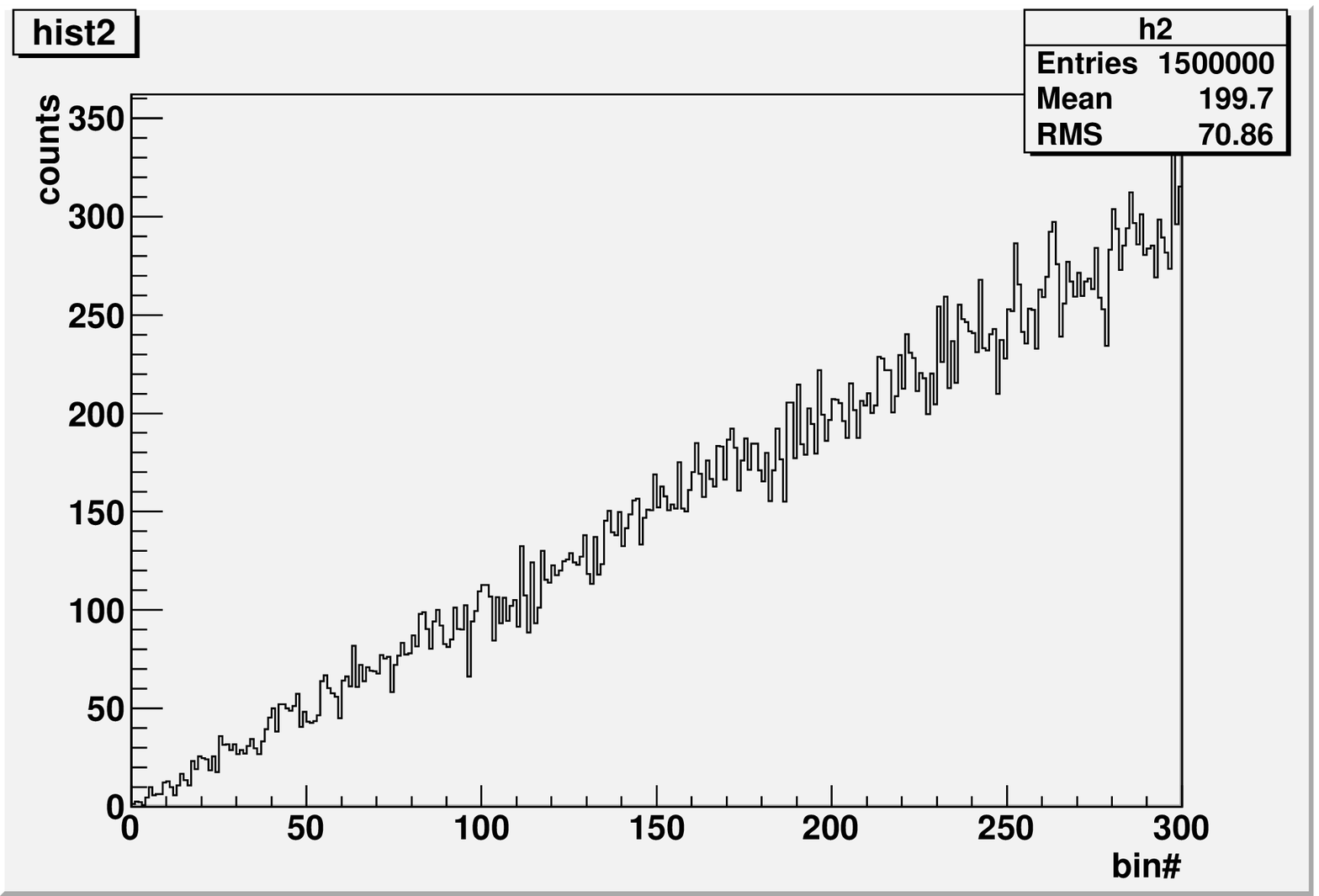}} 
           \resizebox{8.2cm}{!}{\includegraphics{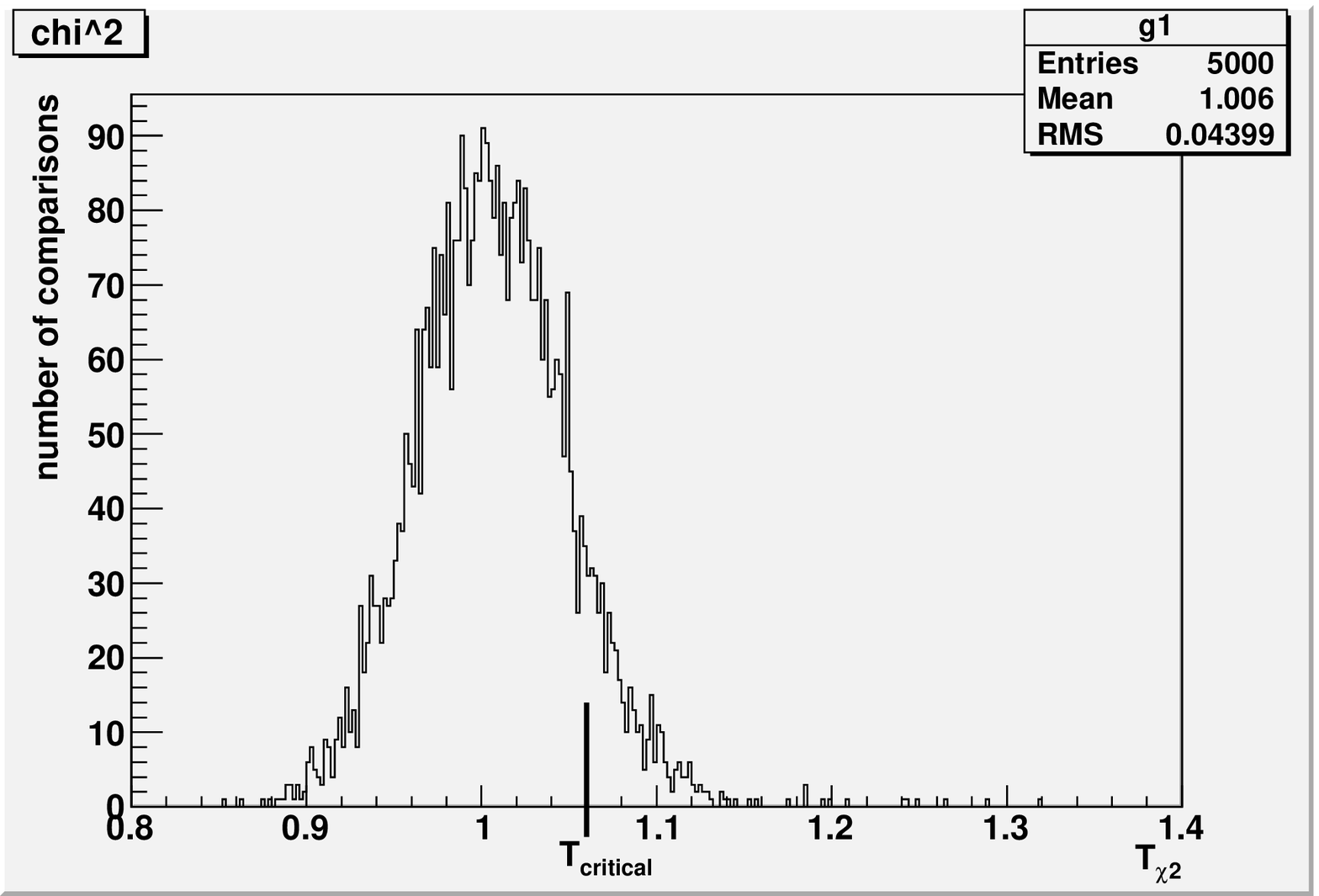}} 
          \resizebox{8.2cm}{!}{\includegraphics{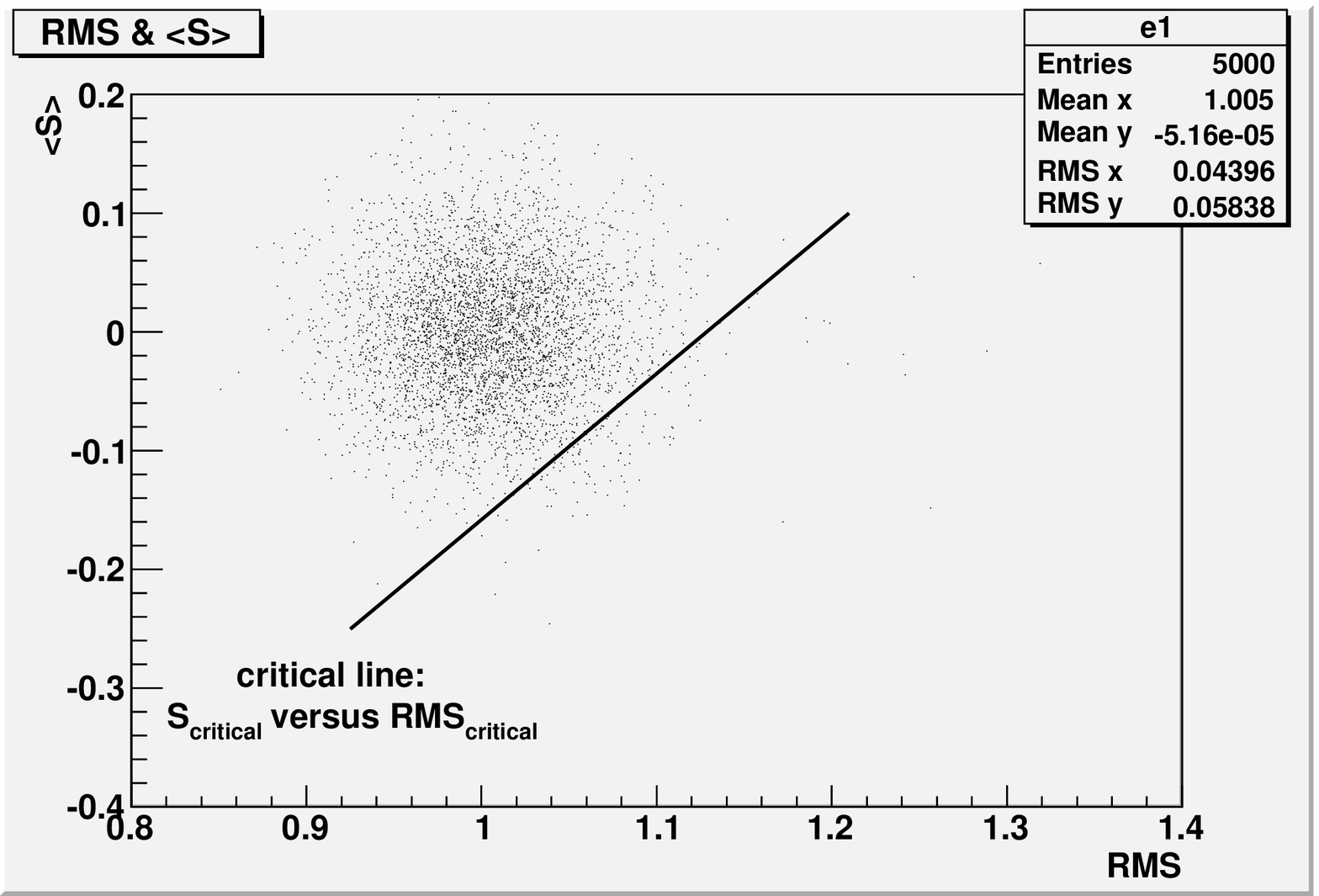}} 
\caption{Case A: Upper histograms are input histograms (triangle distributions, M=300, K=1). 
Down histograms are $T_{\chi^2}$ (left) and $RMS$ \& $\bar S$ (right)  
of the distribution of significances  for 5000 comparisons for input histograms and their clones.}
    \label{fig:2} 
\end{center}
\end{figure}

At first we consider the Case A (Fig.~\ref{fig:2}) when both histograms (hist1 and hist2) 
are obtained from the same statistical population. The distributions of test statistic  
$T_{\chi^2}=\sqrt{\frac{\hat \chi^2}{M}}$ and test statistic $RMS$ versus $\bar S$ 
are produced during 5000 comparisons of histograms (by the use of rehistogramming).

\begin{figure}
\begin{center}
           \resizebox{8.2cm}{!}{\includegraphics{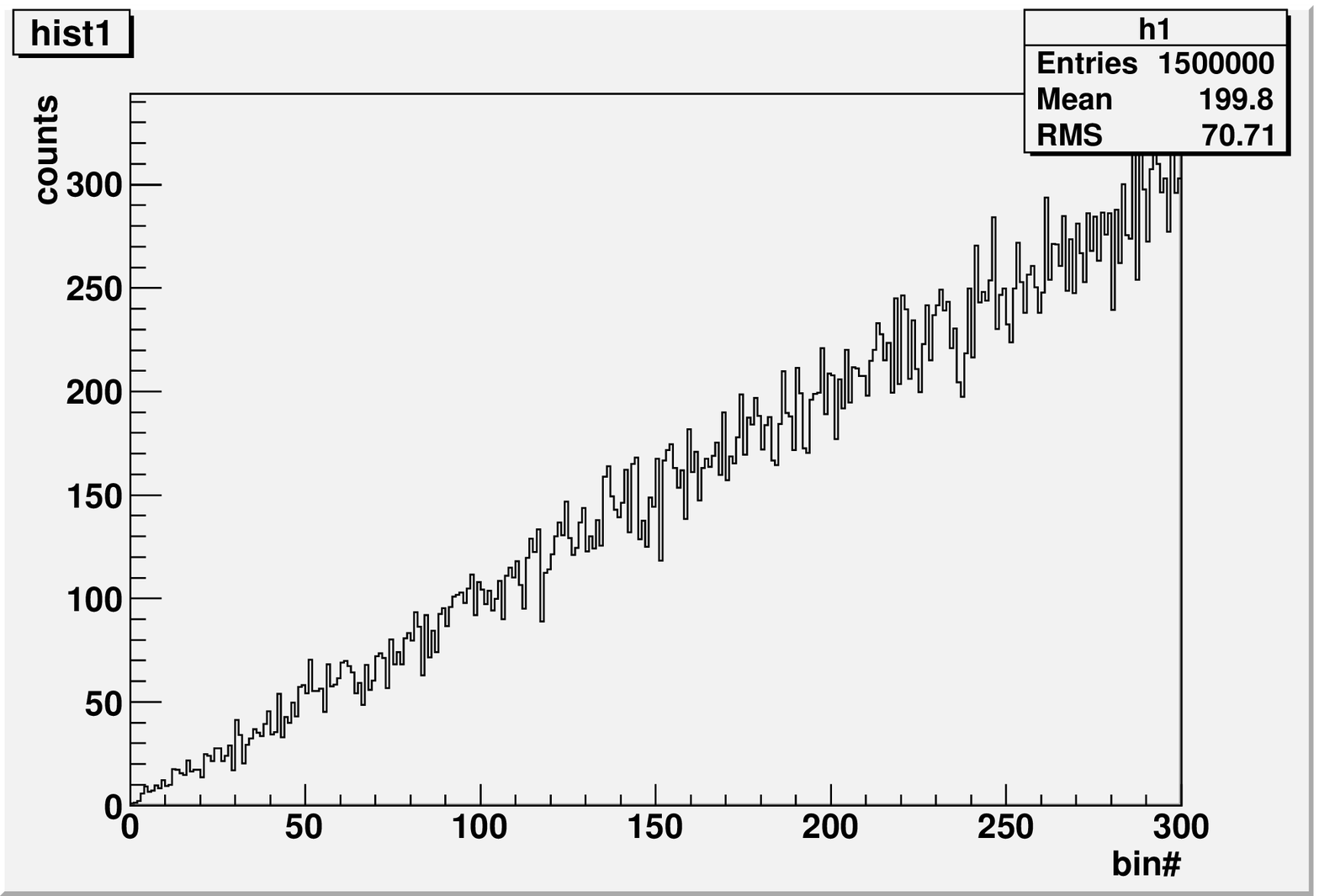}} 
           \resizebox{8.2cm}{!}{\includegraphics{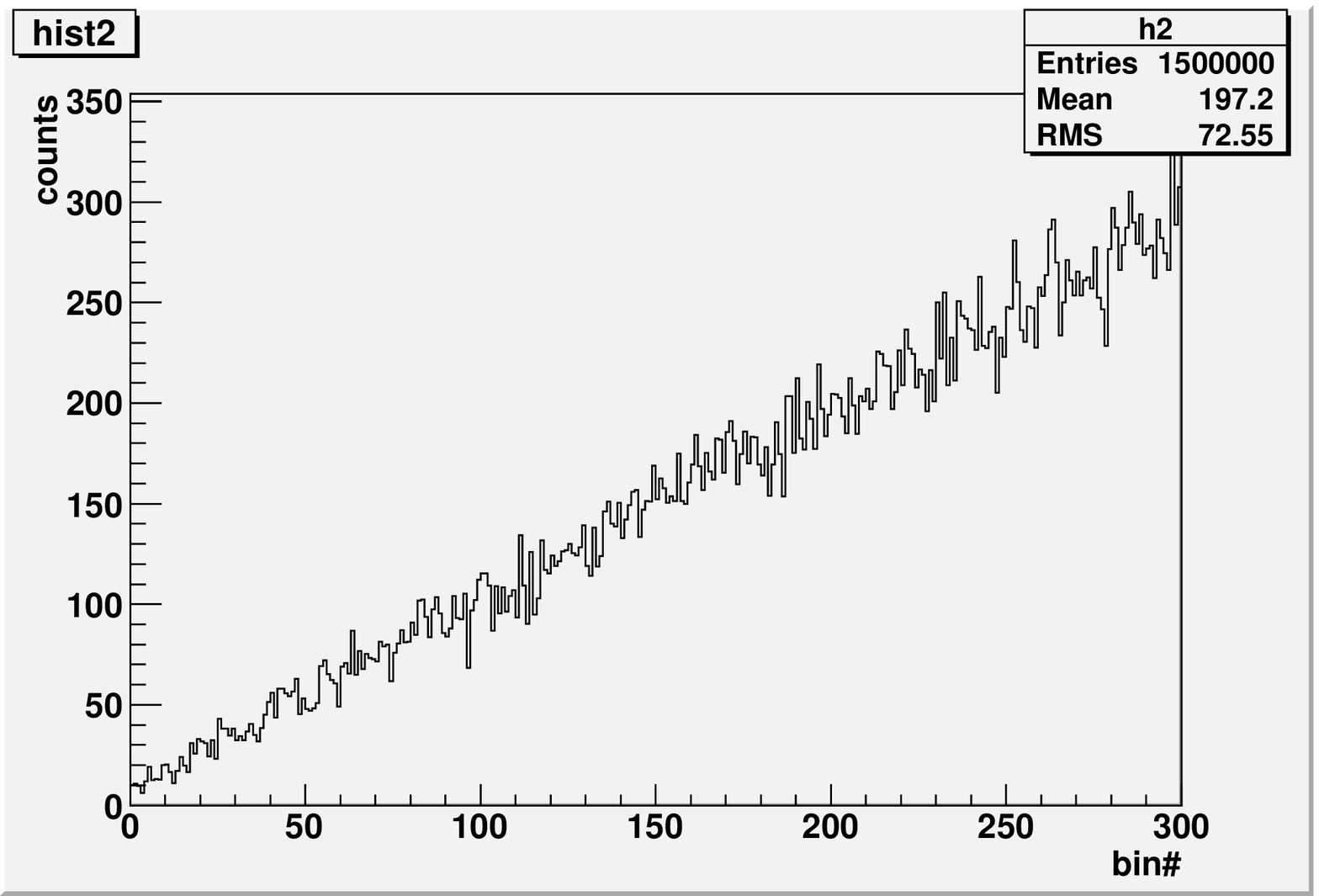}} 
           \resizebox{8.2cm}{!}{\includegraphics{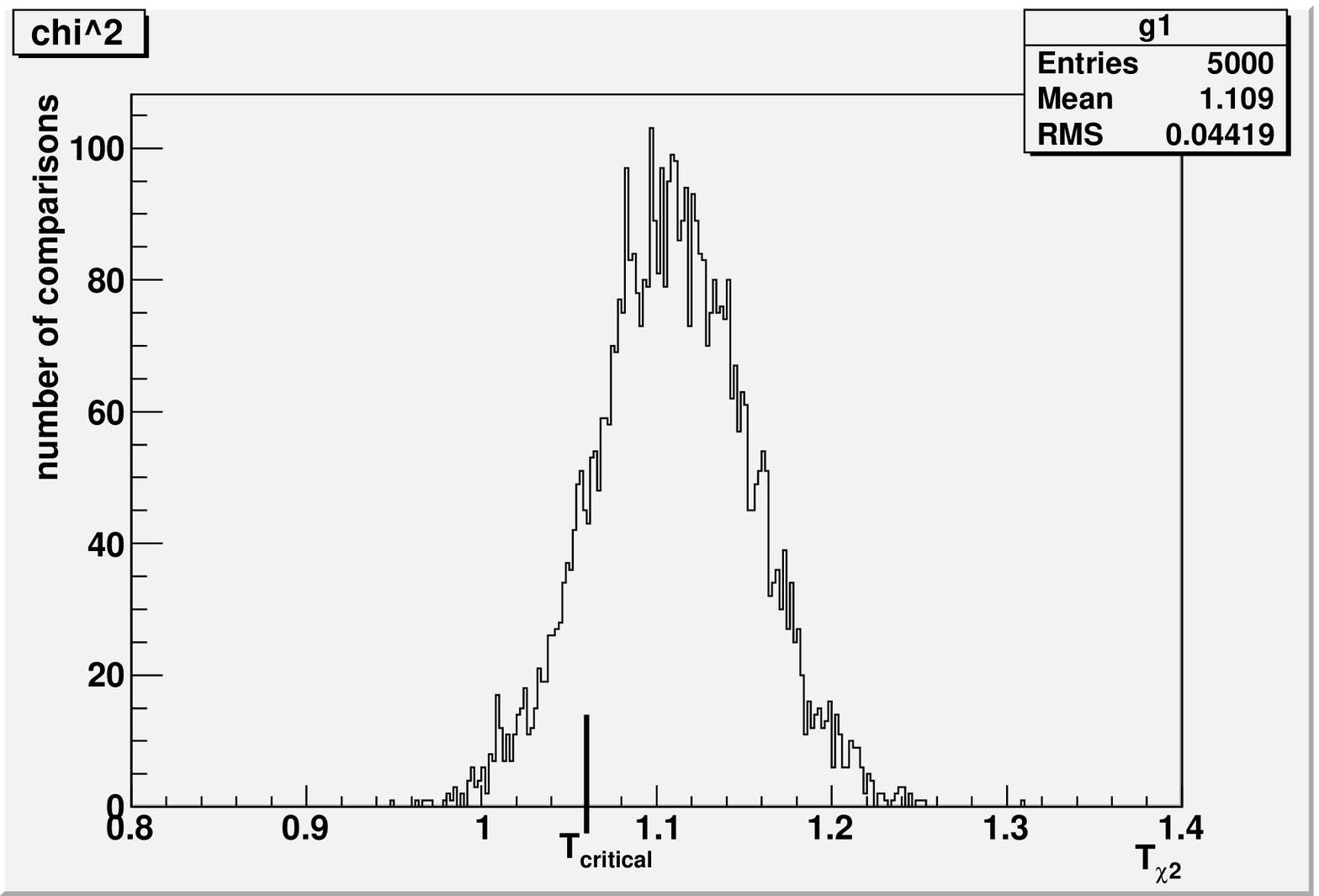}} 
          \resizebox{8.2cm}{!}{\includegraphics{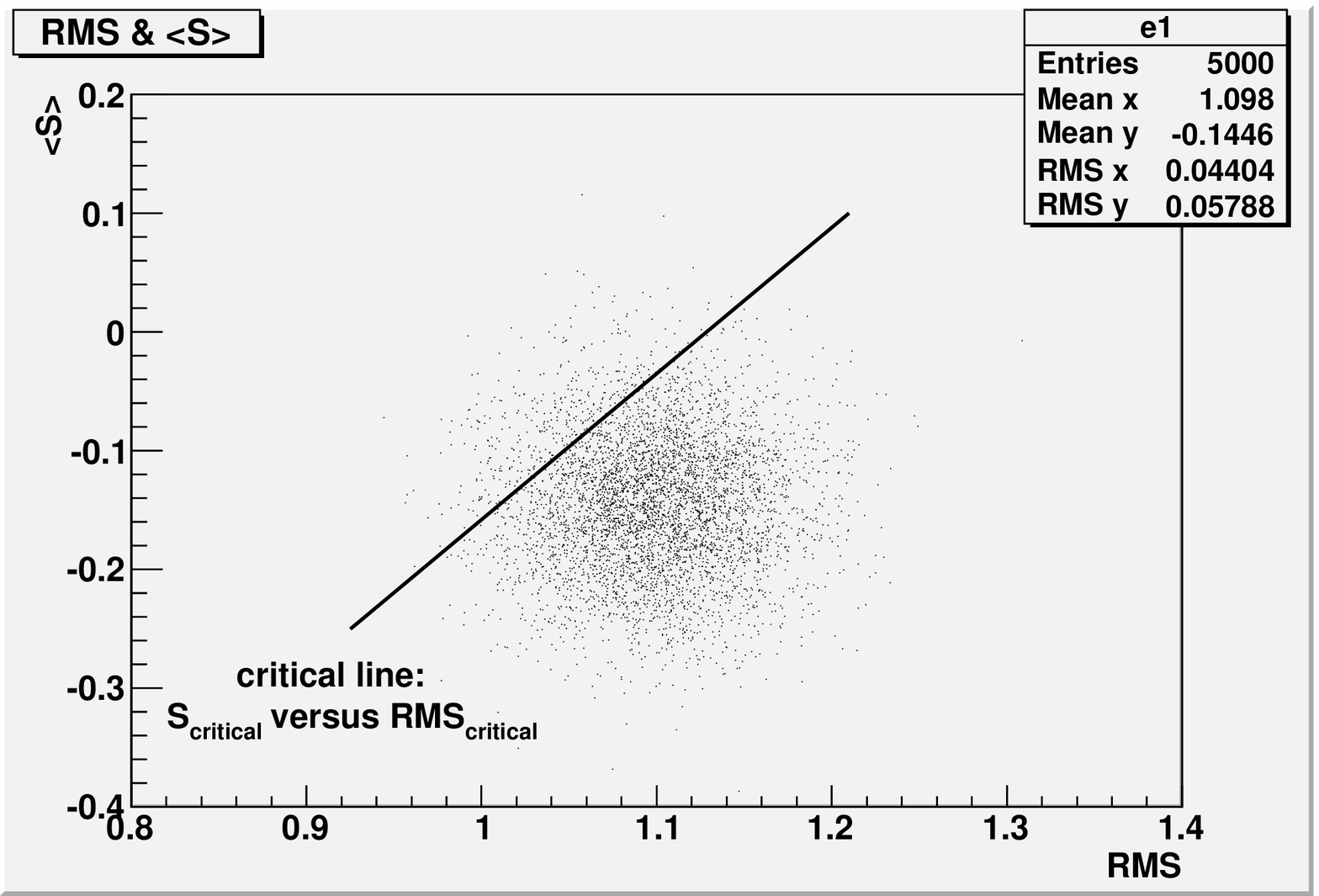}} 
\caption{Case B: Upper histograms are input histograms (the triangle distribution (left) and 
the trapezoidal distribution (right), M=300, K=1). 
Down histograms are $T_{\chi^2}$ (left) and $RMS$ \& $\bar S$ (right) of the distribution of significances 
for 5000 comparisons for input histograms and their clones.}
    \label{fig:3} 
\end{center}
\end{figure}

After that, the content of second histogram (hist2) was changed (Case B), namely, the expected 
content of left bin of histogram was increased from $n_{12}=1.0$ up to $n_{12}=8.5$, 
the expected content of right bin of histogram was decreased from $n_{M2}=300.0$ up to $n_{M2}=292.5$, 
the expected content of other bins was changed to conserve linear dependence between contents in bins. 
The result of the rehistogramming for the Case B is shown in Fig.~\ref{fig:3}. 
One can see that distributions of test statistic $T_{\chi^2}=\sqrt{\frac{\hat \chi^2}{M}}$ and 
test statistic $RMS$ versus $\bar S$ are shifted. 

The probability of  correct decision as a measure for 
distinguishability of two histograms is determined by the 
comparison of distributions for the Case A and corresponding distributions 
for the Case B. The critical value $T_{critical}=1.06$ is used for comparison of 
one-dimensional $T_{\chi^2}$ distributions. The critical line 
$(S_{critical}=1.2\cdot RMS_{critical}-1.36)$ is used for comparison of 
two-dimensional $RMS\&\bar S$ distributions. The results are presented in 
Tab.~\ref{tab:1}. 

For $\chi^2$ method the probability of the correct decision 
($1-\kappa$)  about the Case realization (A or B) is equal to  87.26\%.
For the other method the probability of the correct decision 
($1-\kappa$)  about the Case realization (A or B) is equal to 93.88\%.
One can see that the method, which uses $RMS$ and $\bar S$, 
gives better distinguishability of histograms than the $\chi^2$ method. 
Note that  we  use only two moments of the  significance distributions 
(the first initial moment ($\bar S$) and the square root from the second central 
moment ($RMS$)) for the estimation of distinguishability of histograms. 

\section{Conclusions}

The considered approach allows to perform the comparison of histograms 
in more details than methods which use only one test statistics.
Our  method can be used in tasks of monitoring of the equipment 
during experiments.     

The main items of the consideration are  
\begin{itemize}
\item the normalized significance of deviation provides us the distribution
which is close to $\cal N$(0,1) if $G_1=G_2$;
\item the rehistogramming provides us the tool for an estimation of the accuracy in 
the determination of statistical moments and, correspondingly, for testing  the hypothesis 
about distinguishability of histograms; 
\item the probability of correct decision gives  us the estimator of the decision quality.  
\end{itemize} 

\begin{acknowledgement}

The authors are grateful to L.~Demortier, T.~Dorigo, L.V.~Dudko, V.A.~Kachanov, 
L. Lyons, V.A.~Matveev, L.~Moneta and E. Offermann for the interest and useful comments. 
The authors would like to thank V.~Anikeev, Yu.~Gouz, E.~Gushchin, A. Karavdina, 
D.~Konstantinov, N.~Minaev, A.~Popov, V.~Romanovskiy, S.~Sadovsky and N.~Tsirova for fruitful discussions. 
This work is supported by RFBR grant N 13-02-00363.  

\end{acknowledgement}


{\begin{table}[htb]
\begin{center}
\begin{tabular}{|r|rr||r|rr|} 
\hline
 \multicolumn{3}{|c||}{Distribution of $T_{\chi^2}$} & \multicolumn{3}{c|}{Distribution  of $RMS\&\bar S$} \\  
\hline
         & \multicolumn{2}{c||}{In reality} &  & \multicolumn{2}{c|}{In reality} \\  
Accepted & Case A &  Case B &  Accepted & Case A &  Case B  \\
\hline
Case A   &   4543 &  673    &  Case A   &   4843 &  456      \\
Case B   &    457 & 4327    &  Case B   &    121 & 4544      \\
\hline
\hline
 $1-\kappa$ & $\alpha$ & $\beta$ &  $1-\kappa$ & $\alpha$ & $\beta$  \\
\hline
0.8726      & 0.0914   &  0.1346 &  0.9388     & 0.0242   &  0.0912  \\
\hline
\end{tabular}
\caption{The quality of the hypothesis testing about distinguishability of 
two different histograms for two methods of comparison histograms.}
\label{tab:1}
\end{center}
\end{table}
}


\begin{thebibliography}{99}


\bibitem{Porter} F. Porter, {\it Testing Consistency of Two Histograms}, arXiv:0804.0380.


\bibitem{Gagun} N.D.~Gagunashvili, 
{\it Chi-square tests for comparing weighted histograms}, 
Nucl.Instr.\&Meth., {\bf A614} (2010) 287-296; arXiv:0905.4221.     

\bibitem{arXive}  S.I.~Bityukov, N.V.~Krasnikov, A.N.~Nikitenko, V.V.~Smirnova, 
{\it A method for statistical comparison of histograms}, 
arXiv:1302.2651 [physics.data-an], 2013.   

%

\bibitem{Efron} B. Efron, 
{\it Bootstrap methods: another look at the jackknife}, 
Annals of Statistics, {\bf 7} (1979) 1-26.

\bibitem{NIM534} 
S.I.~Bityukov, N.V.~Krasnikov,    
{\it Distinguishability of Hypotheses},  
Nucl.Inst.\&Meth. A {\bf 534} (2004) 152-155.

\bibitem{ROOT}  R.~Brun, F.~Rademaker, 
{\it ROOT -- An object oriented data analysis framework}, 
Nucl.Instr.\&Meth., A {\bf 389} (1997) 81-86.



     
\end{thebibliography}
\end{document}